\documentclass[12pt,letterpaper]{article}
\usepackage{amsmath,amssymb,array,calc,rotating,epsfig,psfrag, amscd}
\usepackage[left=2cm,top=1cm,right=3cm,nohead]{geometry}
  \newcommand{\nc}{\newcommand}

\nc{\ra}{\rightarrow} 
\nc{\lra}{\leftrightarrow} 
\nc{\Ra}{\Rightarrow} 
\nc{\LRa}{\Leftightarrow} 
\nc{\blp}{{\big (}}
\nc{\brp}{{\big )}}
\nc{\Blp}{{\Big (}}
\nc{\Brp}{{\Big )}}
\nc{\bglp}{{\bigg (}}
\nc{\bgrp}{{\bigg )}}
\nc{\Bglp}{{\Bigg (}}
\nc{\Bgrp}{{\Bigg )}}
\nc{\slb}{{\rm [}}
\nc{\srb}{{\rm ]}}
\def\al{\alpha}

\def\eps{\epsilon}
\nc{\veps}{\varepsilon}
\def\gam{\gamma}

\def\lam{\lambda}
\def\om{\omega}
\nc{\vphi}{\varphi}
\def\tha{\theta}
\def\sig{\sigma}

\def\Gam{\Gamma}

\def\Om{\Omega}
\def\Sig{\Sigma}

\nc{\bea}{\begin{eqnarray}}
\nc{\eea}{\end{eqnarray}}
\nc{\be}{\begin{equation}}
\nc{\ee}{\end{equation}}
\nc{\cA}{{\cal A}}
\nc{\cB}{ \cal B}

\nc{\cF}{{\cal F}}
\nc{\cG}{{\cal G}}

\def\cH{{\cal H}}

\nc{\cL}{{\cal L}}
\nc{\M}{{\cal M}}
\nc{\cM}{{\cal M}}
\def\N{{\cal N}}

\def\cP{{\cal P}}
\nc{\cQ}{{\cal Q}}
\nc{\cR}{{\cal R}}

\def\T{{\cal T}}

\nc{\BB}{{\mathbb B}}
\nc{\CC}{{\mathbb C}}
\nc{\DD}{{\mathbb D}}
\nc{\EE}{{\mathbb E}}
\nc{\FF}{{\mathbb F}}
\nc{\GG}{{\mathbb G}}
\nc{\HH}{{\mathbb H}}
\nc{\JJ}{{\mathbb J}}
\nc{\RR}{{\mathbb R}}
\nc{\PP}{{\mathbb P}}
\nc{\QQ}{{\mathbb Q}}
\nc{\ZZ}{{\mathbb Z}}
\nc{\calone}{{\mathbb 1}}
\nc{\half}{\frac{1}{2}}
\nc{\qrt}{\frac{1}{4}}
\nc{\del}{\partial}

\nc{\delbar}{\bar\partial}
\nc{\Spin}{\operatorname{Spin}}
\nc{\SO}{\operatorname{SO}}

\nc{\Sp}{{\rm Sp}}
\nc{\com}[2]{{ \left[ #1, #2 \right] }}
\nc{\acom}[2]{{ \left\{ #1, #2 \right\} }}
\nc{\rr}{\rightarrow}
\nc{\p}{\partial}
\nc{\LT}{{\LL_\T}}
\nc{\Tr}{{\rm Tr}}
\nc{\tr}{{\rm tr}}
\def\com#1#2{{ \left[ #1, #2 \right] }}
\def\acom#1#2{{ \left\{ #1, #2 \right\} }}
\nc{\ttha}{\tilde{\theta}}
\nc{\tphi}{\tilde{\phi}}
\nc{\tsig}{\tilde{\sig}}
\nc{\tom}{\tilde{\om}}
\nc{\tlam}{\tilde{\lam}}
\nc{\tSig}{\widetilde{\Sig}}
\nc{\tPhi}{\tilde{\Phi}}
\nc{\tPi}{\tilde{\Pi}}
\nc{\tpsi}{\tilde{\psi}}
\nc{\tPsi}{\tilde{\Psi}}
\nc{\tgam}{\tilde{\gam}}
\nc{\tGam}{\tilde{\Gam}}
\nc{\tb}{\tilde b}
\nc{\tc}{\tilde c}
\nc{\te}{\tilde e}
\nc{\tf}{\tilde f}
\nc{\tg}{\tilde g}
\nc{\tj}{\tilde j}
\nc{\tp}{\widetilde{p}}
\nc{\tq}{\widetilde{q}}
\nc{\ts}{{\tilde s}}
\nc{\tz}{\tilde z}
\nc{\tA}{{\tilde A}}
\nc{\tAbar}{{\ol \tA}}
\nc{\tD}{{\tilde D}}
\nc{\tE}{{\tilde E}}
\nc{\tG}{{\tilde G}}
\nc{\tH}{{\tilde H}}
\nc{\tJ}{{\tilde J}}
\nc{\tJbar}{{\ol {\tilde J}}}
\nc{\tM}{{\tilde M}}
\nc{\tN}{{\tilde N}}
\nc{\tP}{{\tilde P}}
\nc{\tQ}{{\tilde Q}}
\nc{\tS}{\tilde{S}}
\nc{\tF}{\tilde{{\cal F}}}
\nc{\tX}{\widetilde{X}}
\nc{\hb}{\hat b}
\nc{\hc}{\hat c}
\nc{\hd}{\hat d}
\nc{\he}{\hat e}
\nc{\hf}{\hat f}
\nc{\hg}{\hat g}
\nc{\hh}{\hat h}
\nc{\hp}{\hat p}
\nc{\hv}{\hat v}
\nc{\hw}{\hat w}
\nc{\hx}{\hat x}
\nc{\hy}{\hat y}
\nc{\hz}{\hat z}
\nc{\hA}{\widehat{A}}
\nc{\hE}{\widehat{E}}
\nc{\hF}{\widehat{F}}
\nc{\hH}{\widehat{H}}
\nc{\hJ}{\widehat{J}}
\nc{\tK}{\widetilde{K}}
\nc{\hM}{\widehat M}

\nc{\ha}{\widehat \alpha}
\nc{\hphi}{\hat{\phi}}
\nc{\hpsi}{\hat{\psi}}
\nc{\hgam}{\hat{\gam}}
\nc{\hPhi}{\hat{\Phi}}
\nc{\hPsi}{\hat{\Psi}}
\nc{\hGam}{\hat{\Gam}}

\nc{\w}{\wedge}

\nc{\ol}{\overline}
\nc{\abar}{\ol{a}}
\nc{\bbar}{\ol{b}}
\nc{\cbar}{\ol{c}}
\nc{\dbar}{\ol{d}}
\nc{\ebar}{\ol{e}}
\nc{\ibar}{\ol{\imath}}
\nc{\jbar}{\ol{\jmath}}
\nc{\kbar}{\ol{k}}
\nc{\lbar}{\ol{l}}
\nc{\mbar}{\ol{m}}
\nc{\nbar}{\ol{n}}
\nc{\pbar}{\ol{p}}
\nc{\qbar}{\ol{q}}
\nc{\ubar}{\ol{u}}
\nc{\vbar}{\ol{v}}
\nc{\wbar}{\ol{w}}
\nc{\xbar}{\ol{x}}
\nc{\ybar}{\ol{y}}
\nc{\zbar}{\ol{z}}

\nc{\Abar}{\ol{A}}
\nc{\Dbar}{\ol{D}}
\nc{\Ebar}{\ol{E}}
\nc{\Jbar}{\ol{J}}
\nc{\Kbar}{\ol{K}}
\nc{\Lbar}{\ol{L}}
\nc{\Pbar}{\ol{P}}
\nc{\Qbar}{\ol{Q}}
\nc{\Wbar}{\ol{W}}
\nc{\Xbar}{{\overline X}}
\nc{\Ybar}{{\overline Y}}
\nc{\Zbar}{{\overline Z}}

\nc{\epsbar}{\ol{\epsilon}}
\nc{\lambar}{\ol{\lambda}}
\nc{\psibar}{\ol{\psi}}
\nc{\Psibar}{\ol{\Psi}}
\nc{\phibar}{\ol{\phi}}
\nc{\Phibar}{\ol{\Phi}}
\nc{\chibar}{\ol{\chi}}
\nc{\mubar}{\ol{\mu}}
\nc{\nubar}{\ol{\nu}}
\nc{\rhobar}{\ol{\rho}}
\nc{\ombar}{\ol{\om}}
\nc{\Ombar}{\ol{\Om}}
\nc{\sinp}{s_{\phi}}
\nc{\cosp}{c_{\phi}}
\nc{\tanp}{t_{\phi}}
\nc{\spone}{s_{\phi_1}}
\nc{\cpone}{c_{\phi_1}}
\nc{\tpone}{t_{\phi_1}}
\nc{\sptwo}{s_{\phi_2}}
\nc{\cptwo}{c_{\phi_2}}
\nc{\tptwo}{t_{\phi_2}}
\nc{\spth}{s_{\phi_3}}
\nc{\cpth}{c_{\phi_3}}
\nc{\tpth}{t_{\phi_3}}

\nc{\bah}{{\mathbf {\hat{A}}}}
\nc{\bX}{{\mathbf X}}
\nc{\bj}{{\bf j}}
\nc{\bk}{{\bf k}}
\nc{\bu}{{\bf u}}
\nc{\bv}{{\bf v}}
\nc{\bom}{{\bf \om}}
\nc{\bombar}{{\mathbf \ombar}}
\nc{\dal}{\dot{\al}}
\nc{\thab}{\bar{\theta}}
\nc{\thal}{\theta^{\al}}
\nc{\thdal}{\bar{\theta}^{\dal}}

\nc{\thsigthm}{\tha \sigma^m \thab}
\nc{\thsigthn}{\tha \sigma^n \thab}

\nc{\Dal}{D_{\al}}
\nc{\Ddal}{\bar{D}_{\dal}}
\nc{\CDal}{{\cal D}_{\al}}
\nc{\CDdal}{\bar{\cal D}_{\dal}}

\nc{\eq}[1]{(\ref{#1})}
\nc{\non}{\nonumber}
\nc{\equ}{{\rm eq}}

\nc{\AdS}{{\rm AdS}}
\nc{\vol}{{\rm vol}}
\nc{\Ainf}{A_{\infty}}
\nc{\End}{{\rm End}}
\nc{\Ext}{{\rm Ext}}
\nc{\Hom}{{\rm Hom}}
\nc{\IIB}{{\rm IIB}}
\nc{\Dslash}{\ensuremath \raisebox{0.025cm}{\slash}\hspace{-0.32cm} D}
\nc{\no}{\!:\!\!}
\nc{\ointdz}{\oint\frac{dz}{2\pi i}}
\nc{\ointdzone}{\oint\frac{dz_1}{2\pi i}}
\nc{\ointdztwo}{\oint\frac{dz_2}{2\pi i}}
\nc{\ointdzb}{\oint\frac{d\zbar}{2\pi i}}
\nc{\ointdzbone}{\oint\frac{d\zbar_1}{2\pi i}}
\nc{\ointdzbtwo}{\oint\frac{d\zbar_2}{2\pi i}}
\nc{\dz}{\frac{dz}{2\pi i}}
\nc{\dzb}{\frac{d\zbar}{2\pi i}}
\nc{\bpm}{\begin{pmatrix}}
\nc{\epm}{\end{pmatrix}}
 \nc{\bitem}{\begin{itemize}}
 \nc{\eitem}{\end{itemize}}

\begin{document}
\begin{center}
\today \hfill     EFI-08-14
\vskip 2 cm

{\Large \bf  Non-geometric String Backgrounds\\ \vskip 3mm and  Worldsheet Algebras}\\

\vskip 1.25 cm 
{ Nick Halmagyi\footnote{email address: halmagyi@theory.uchicago.edu} }\\
{ \vskip 0.5cm Enrico Fermi Institute, \\
University of Chicago, \\Chicago, IL 60637, USA\\}
\vskip 1cm
\end{center}
\vskip 1 cm

\begin{abstract}
Using worldsheet Hamiltonian methods we derive a charge algebra which generalizes the Courant bracket to include fluxes of general index type. This is achieved by coupling a bi-vector to the Hamiltonian of the Polyakov model. This bracket is useful to describe so-called {\it non-geometric} backgrounds and has been discussed in the mathematics literature by Dmitry Roytenberg. 
\end{abstract}
%%%%%%%%%%%%%%%%%%%%%%%%%%%%%%%%%%%%

%%%%%%%%%%%%%%%
%%%%%%%%%%%%%%%

\section{Introduction}

The motivation for the current work comes from a class of string backgrounds which have been known as {\it non-geometric backgrounds} since the pioneering work \cite{Hellerman:2002ax}. This moniker comes from the fact that unlike manifolds, which are sewn together patch to patch by diffeomorphisms, these backgrounds allow for the possibility that they are sewn together using more general elements of the stringy duality group. On circle bundles over circles this duality group is $O(d,d)$. 

It is well known that the metric and fluxes of string theory should be treated somewhat equally and to this effect string duality allows for these to be transformed into each other. Indeed the duality group of compactified string theory is enlarged significantly by the spacetime fluxes. The central tenet of the non-geometric theme advocated in \cite{Hellerman:2002ax} is that since these enlarged duality groups include and extend diffeomorphisms, one should allow for string backgrounds which are locally geometric but can be patched together by  elements of this enlarged duality group.

Various examples have by now been put forward, in the original work \cite{Hellerman:2002ax} a non-geometric version of $K3$ was studied with the SYZ $T^2$ fibration used to locally model a torus compactification. There is a somewhat simpler example of a $T^2$ fibered over $S^1$ \cite{Flournoy:2004vn, Flournoy:2005xe, Hellerman:2006tx} and in fact in the latter of these papers, the authors have succeeded in finding a modular invariant worldsheet CFT which describes this background.

One more example which unveils some of the relevant structures is to consider a $T^3$ with $N$ units of $H$-flux (from here on this will be referred to as $T3H3$). This is not a conformal background since it clearly does not satisfy Einstein's equation, however it can be rendered conformal by embedding it is the full ten dimensional string theory and this retains the essential features relevant for the current discussion \cite{Ellwood:2006my}. This background appears to admit three isometries however when one utilizes any one of these to perform T-duality, there are apparently then only two isometries of the background. This disagrees with intuition garnered from a mountain of previous work on dualities in string theory. One of the known duality frames is a twisted torus and another is a prime example of a non-geometric background known as the $Q$-space. We will discuss this example at some length in the current work and in fact we use it as inspiration for some formal developments. However the structures developed here are not restricted to this example and are ultimately best thought of as somewhat independent from it.

In this work we study non-geometric structures by coupling a bi-vector to the Polyakov string model. Our main result  is a worldsheet derivation of the Roytenberg bracket. This bracket is a map 
\be
 [.,.]_R:T_X\oplus T^*_X \times T_X\oplus T^*_X\ra T_X\oplus T^*_X
 \ee 
which generalizes the better known Courant bracket. A worldsheet derivation of the Courant bracket  (to be more precise the Dorfman bracket, see section \ref{courant}) has been provided by Alekseev and Strobl \cite{Alekseev:2004np} where they realize this bracket as the current algebra of the non-supersymmetric Polyakov model. This calculation considered the standard Polyakov action and as such had both metric and $B$-field, the key addition to the story provided in this paper is the twisting by a bi-vector. As a result of this twisting, there are extra co-efficients in the resulting algebra of worldsheet charges precisely corresponding to those in the Roytenberg bracket \cite{roytenberg-2002-61}. 

We verify that this bracket correctly describes the various known duality frames of the $T3H3$ example. We also discuss the final duality frame of this example, the conjectural $R$-space but are not able to provide the bi-vector for this background. On the other hand, we cannot rule out the possibility that the $R$-space can be described successfully in this formalism. 

This paper is organized as follows: In section 2 we review some general aspects of non-geometric string backgrounds and the $T3H3$ backgrounds and its sibling duality frames in particular. In section 3 we review the worldsheet derivation of the Courant bracket. Section 4 contains the main result of the paper, a worldsheet derivation of the Roytenberg bracket. Section 5 is devoted to fleshing out the $T3H3$ duality frames in explicit detail to demonstrate the utility of the current approach. The appendix contains a brief exposition on the WZ-Poisson sigma model.

%%%%%%%%%%%%%%%%%%%%%%%%%%%%%%%%%%%%

\section{Torus with $H$-Flux and T-duality} \label{dualitygen}

We will warm up by considering the $T3H3$ background  and its duality frames. Two somewhat separate issues occur in this context, firstly we encounter shortcomings of the local T-duality rules of Buscher \cite{Buscher:1987qj} and secondly we find in a certain duality frame (the $Q$ space) a simple example of a non-geometric background. Global aspects of T-duality and this duality sequence in particular have been studied in several more mathematical papers \cite{Bouwknegt:2004ap, Mathai:2004qq, Mathai:2004qc,Belov:2007qj}.

The problem first appears in the $T3H3$ background when one chooses a gauge for the B-field, 
\bea
ds^2&=&dw^2+dx^2+dy^2, \non \\
B&=&wdx\w dy, \non \\
\Rightarrow H&=&dw\w dx\w dy \non.
\eea
At this point, the background exhibits only two symmetries (translation in the $x$ and $y$ directions) although we know that the gauge invariant couplings come from the $H$-field which couples to the string worldsheet via a WZW term.  So we can surmise that in the fiull string theory there are actually three symmetries.

After T-dualizing along the $y$ direction one obtains the purely metric background (called the {\it twisted torus})
\bea
ds^2 &=& dw^2+ dx^2 + (dy-Nw dx)^2 \non \\
B&=&0.
\eea
In this frame basis there in a non-trivial bracket on the vector fields and one interprets this by saying that  the integral $H$-flux quanta have been transformed into {\it geometric flux}. To see all components of the structure constants one has to use the Courant bracket (see section \ref{T3H3})

One can perform a further T-duality on the twisted torus obtaining the background
\bea
ds^2&=&dw^2+\frac{1}{1+N^2 w^2}(dx^2 + dy^2), \non \\
B&=&\frac{Nw}{1+N^2 w^2} dx\w dy. \label{Qspace}
\eea
At first order in an expansion in $N$  this is the original $T3H3$ background however with the $N$-dependent corrections, the space is globally not a manifold. This space can be thought of a $T^2$ fibration over $S^1$ where upon winding around the base $S^1$ the complexified Kahler form  ($\rho= B+i J$) of the fiber shifts as
\be
\frac{1}{\rho} \ra \frac{1}{\rho} + N.
\ee
As a result the transitions functions mix the $B$-field and the metric. In addition the volume shifts as $w\ra w+1$ so regions where the volume approaches the string scale and there the background receives large stringy corrections.

A useful notation was introduced in \cite{Shelton:2005cf} in which this background is referred to as the $Q$-space. The basic idea is as follows: T-duality exchanges winding and momentum and as such swaps an upper index for a lower one. So in this chain of dualities the quantized charges transform as
\be
H_{abc}\stackrel{T}{\lra} f^{a}_{\ bc} \stackrel{T}{\lra}  Q^{ab}_{\  c}.
\ee

An important and unresolved question is whether there exists a duality which would produce  a charge with three indices raised
\be
Q^{ab}_{\ \ c}\stackrel{?}{\lra} R^{abc}.
\ee
Although some evidence for this was produced by studying the superpotential of the four dimensional effective supergravity of this background \cite{Shelton:2005cf}, it utilized the Gukov-Vafa-Witten superpotential which a ten dimensional gravity expression. As mentioned above gravity is not a good approximation for the $Q$-space. We will argue in a later section that while this is problematic, it is still not quite the essence of the paradox.

Regardless of whether this final duality is in fact valid, one can consider the possibility of backgrounds with general charges $H,f,Q,R$ although no explicit examples are currently known. The present work will demonstrate that by considering backgrounds in the presence of a bi-vector, one can in principle produce the above charges.

%%%%%%%%%%%%%%%%%%%%%%%%%%%%%%%%%%%%
%%%%%%%%%%%%%%%%%%%%%%%%%%%%%%%%%%%%

\section{The Courant Bracket} \label{courant}

The Courant bracket is a natural generalization of the Lie bracket on vector fields,
taking as its arguments sections of $T_X\oplus T^*_X$. A clear exposition is given in section 3 of Gualtieri's thesis \cite{Gualtieri:2003dx}. It has appeared in the physics literature through generalized complex geometry, where closure under the Courant bracket implies integrability of the generalized complex structure. 

The co-ordinate independent expression for the Courant bracket is:

\be
[u+\alpha, v+\beta]_C = [u,v] + \cL_{u}\beta - \cL_{v} \alpha - \half d(\iota_{u} \beta - \iota_v \alpha) + H(u,v,.)\label{Courant}
\ee
where $u,v\in \Gamma(T_X)$ and $\alpha, \beta \in \Gamma(T^*_X)$. The Courant bracket is skew symmetric but does not necessarily satisfy the Jacobi identity, however the Jacobi identity will be satisfied if $T_X$ and $T^*_X$ are {\it Dirac structures}. There is another bracket on $T_X\oplus T_X^*$ called the Dorfman bracket which is occasionally confused with the Courant Bracket
\be
[u+\alpha, v+\beta]_D = [u,v] + \cL_{u}\beta - \iota_{v}d \alpha + H(u,v,.) .
\ee
The Courant bracket is the skew-symmetrization of the Dorfman bracket. In fact the Courant bracket and Dorfman bracket differ by a total derivative
\be
[u+\alpha, v+\beta]_C=[u+\alpha, v+\beta]_D-d\langle(u,\alpha), (v,\beta)\rangle \label{CD}
\ee
where $\langle.,.\rangle$ is the symmetric nondegenerate bilinear form on $T_X\oplus T^*_X$.  When evaulated on Dirac structures, the two brackets are equal since the condition for a Dirac structure is $\langle(u,\alpha), (v,\beta)\rangle=0$.

In a nice paper by Alekseev and Strobl \cite{Alekseev:2004np}, they showed how to derive the Dorfman bracket from a worldsheet current algebra calculation. In fact to get the exact result they required the currents to be formed from Dirac structures. As we will show below, by computing instead the algebra of charges, the distinction between the Courant and Dorfman brackets is lost since by \eq{CD} they differ by total derivative and so the algebra of charges is exactly the Courant bracket.

From a string theory point of view, the key feature of the Courant bracket is that its group of automorphisms is a semi-direct product of diffeomorphisms and shifts by closed $B$-fields. 

%%%%%%%%%%%%%%%%%%%%%%%%%%%%%%%%%%%%

\subsection{From the Worldsheet} \label{courantws}

The Hamiltonian for the Polyakov model is
\be
\cH_P= \half\int d\sig\Blp  G^{ij} (p_i+ B_{ik} \del x^k) (p_j+B_{jl} \del x^l) + G_{ij} \del x^i \del x^j \Brp \label{PolyakovH}
\ee
and with this form of the Hamiltonian the symplectic form is in the Darboux form
\be
\om= \int d\sig \delta x^i \w \delta p_i.
\ee
It is convenient to change co-ordinates on phase space 
\be
p_i\ra p_i - B_{ij} \del_\sig x^j \label{ptransform}
\ee
which results in the following Poisson brackets (see Appendix \ref{Ap1}):
\bea
\{ x^i(\sig),x^j(\sig')\}&=&0 ,\non \\
 \{x^i(\sig) , p_j(\sig')\}&=&\delta^i_j \delta(\sig-\sig'),\label{twpoisson} \\
\{p_i(\sig),p_j(\sig') \} &=&-H_{ijk}\del x^k \delta (\sig-\sig').  \non
\eea

In these co-ordinates, the most general currents are
\be
J_{(u,\alpha)} = u^i p_i  + \alpha_j \del_\sig x^j \label{J1}
\ee
and of course the associated charges are given by $\cQ_{(u,\alpha)}=\int d\sig J_{(u,\alpha)}(\sig)$. The first term in \eq{J1} is a vector field on $X$ and the second term a one form on $X$.  The charge algebra is then found to be

\bea
\{ \cQ_{(u,\alpha)},\cQ_{(v,\beta)}\}&=&\int d\sig d \sig'\{( u^i p_i  + \alpha_j \del_\sig x^j) (\sig),(v^i p_i  + \beta_j \del_{\sig'} x^j)(\sig') \} \non \\
&=& -\int d\sig \Blp [u,v]^ip_i  + ( u^k \del_k \beta_j -v^k \del_k \alpha_j +u^k v^l H_{klj})\del x^j \Brp (\sig) \non \\
&&- \int d\sig d \sig' \Blp u^i(\sig)\beta_i(\sig') \del_{\sig'}\delta(\sig-\sig')-v^i(\sig') \alpha_i(\sig))\del_{\sig}\delta (\sig-\sig') \Brp \non \\
&=& -\int d\sig \Blp [u,v]^ip_i  + (( \cL_{\bu} \beta-\cL_{\bv}\alpha - \half d(\iota_{u} \beta - \iota_v \alpha))_j+u^k v^l H_{klj})\del x^j \Brp (\sig) \non \\
&=&-\cQ_{[(u,\alpha),(v,\beta)]_C}
\eea
where we recall now the expression for the Lie derivative 
\be
\cL_{\bu}=\iota_{\bu} d + d\,  \iota_{\bu}.
\ee
It is important that we have included the spatial integration and thus computed the charge algebra not the current algebra, this sets the total derivative terms in \eq{Courant} to zero and as such $\cL_{\bu}\sim \iota_{\bu} d$ and we recall the explicit co-ordinate expression
\be
\iota_{\bu} d \alpha= u^k \del_k \alpha_i dx^i -u^k \del_i \alpha_k dx^i .
\ee 

So we have shown that the Courant bracket arises from a worldsheet charge algebra. The Courant bracket satisfies the Jacobi identity only when evaluated on Dirac structures however since this worldsheet computation is a Poisson bracket, it must always satisfy the Jacobi identity. This conflict is resolved by computing the charge algebra instead of the current algebra and thus total derivatives terms in the bracket vanish.

%%%%%%%%%%%%%%%%%%%%%%%%%%%%%%%%%%%%

\subsection{Conserved Charges} \label{cons1}

Having established a connection to the Courant bracket for these charges it is worth considering when they are actually conserved. One finds that
\bea
\{\cQ_{(u,\alpha)}, \cH \}&=&\int d\sig \cL_{\bu}(G_{ij})\blp G^{ik}G^{jl} p_k p_l  - \del x^i \del x^j \brp(\sig) \non \\
&& + 2 \int d\sig \blp -u^i  H_{ijl} +(\del_j \alpha_l - \del_l \alpha_j ) \brp G^{jk} p_k \del x^l (\sig),   \non 
\eea
so we see that this is conserved if 

\be
\cL_{\bu}G=0,\ \ \ \cL_{\bu}B=d(\alpha-\iota_\bu B). \label{conserved}
\ee
We now see that the very stringy addition of the term $\alpha_i \del_\sig x^i$ to the usual currents in \eq{J1} and thus the appearance of $T_X^*$ is due to the NS $B$-field. 

It should be stressed however that the Courant bracket appears on the worldsheet regardless of whether the currents are conserved. Indeed, in the twisted torus example and also the $Q$-space, we will derive the correct charge algebra but only two of the three currents are conserved.

%%%%%%%%%%%%%%%%%%%%%%%%%%%%%%%%%%%%
%%%%%%%%%%%%%%%%%%%%%%%%%%%%%%%%%%%%

\section{The Roytenberg Bracket}

We now wish to consider the generalization of the Courant bracket to include a bi-vector $\Pi^{ij}$, which we will call the 
Roytenberg bracket. We will generalize the worldsheet current algebra calculation of the previous section to derive this bracket.

%%%%%%%%%%%%%%%%%%%%%%%%%%%%%%%%%%%%

\subsection{Some Brief Mathematical Preliminaries}

The bracket which we are set to study and which will possibly buttress the theory of non-geometric string backgrounds has in fact been discovered some time ago by Dimitry Roytenberg \cite{roytenberg-2002-61}. The Roytenberg bracket is an extension of the Courant bracket to include a bi-vector and also a three vector. 

In a particular basis, the Roytenberg bracket is
\bea
{\rm [}e_a,e_b{\rm ]} &=& f_{ab}^{\ \ c}  e_c + H_{abc} e^c \non \\
{\rm [}e_a,e^b{\rm ]}  &=& Q_{a}^{\ bc} e_c + f_{a\ c}^{\ b} e^c \label{HfQR}\\
{\rm [}e^a,e^b{\rm ]} &=& R^{abc} e_c + Q^{ab}_{\ \ c} e^c \non
\eea
where we have introduced notation which differs from that of Roytenberg but is hopefully clear. One should compare this expression to the formulas at the top of page 4 in the nice review paper \cite{kosmannschwarzbach-2003}. The expressions for the structure constants will be given once we have calculated the bracket from the worldsheet, see eq. \eq{royt}. From a mathematical point of view the Roytenberg bracket gives families of examples of Courant algebroids. 

The Courant bracket is clearly asymmetric with respect to $T_X$ and $T^*_X$ since it is twisted by a three form $H$ but not by a three vector. The Roytenberg bracket rectifies this somewhat although it does not restore the symmetry completely. One step in restoring this symmetry is to introduce a differential \cite{severa-2001-144} which maps $p$-vectors to $(p+1)$-vectors to go with the usual (twisted)-exterior derivative $d_H$
\bea
d_H&=&d+H\w  \\
d_{\Pi}&=&[\Pi,.]_S + \w^2\Pi H ,
\eea 
where  $\Pi$ is a bi-vector and $[.,.]_S$ is the Schouten bracket. The differential $d_{\Pi}$ will square to zero iff the twisted Poisson condition holds \cite{kosmannschwarzbach-2004-69}
\be
\half[\Pi,\Pi]_S =\w^3\Pi H. \label{dpisq}
\ee

A very curious fact is that this condition \eq{dpisq} appears as the component of the $R^{ijk}$ co-efficient in \eq{HfQR} which comes from the bi-vector $\Pi$. In the math literature, they allow for a three vector  in addition  which is not equal to $d_{\Pi} \Pi$. While in this work we will find a concise physical interpretation for the bi-vector $\Pi$, it is less clear physically how to incorporate another arbitrary ($d_{\Pi}$ closed) three-vector.

%%%%%%%%%%%%%%%%%%%%%%%%%%%%%%%%%%%%

\subsection{The Hamiltonian} \label{hamiltonian}

The Polyakov Hamiltonian   \eq{PolyakovH} can be rewritten by trading the two form field $B_{ij}$ for a two vector $\Pi^{ij}$
\be
\cH_{\Pi}= g^{ij} p_i p_j + g_{kl} (\del x^k + \Pi^{ki}p_i )(\del x^l + \Pi^{lj} p_j)  .
\ee
The data $(g,\Pi)$ can be related to the data $(G,B)$ as follows
\bea
G_{ij}&=& (g- \Pi g  \Pi )^{-1}_{ij},  \label{Gg} \\
B_{ij}&=& G_{ik} g_{jl} \Pi^{kl},\non \\
&=& (g- \Pi g  \Pi )^{-1}_{ik} ~ g_{jl} \Pi^{kl}.\label{BPi} 
\eea

Interestingly, one obtains the same formulas again by considering the Lagrangian
\be
\cL_{A}=  g^{ij} A_i \w * A_j + \Pi^{ij} A_i \w A_j + A_i \w dx^i \label{LagA}
\ee
and integrating out the $A_i$ fields to obtain the usual Polyakov Lagrangian
\be
\cL_{\cP}= G_{ij} dx^i \w * dx^j + B_{ij} dx^i \w dx^j. \label{LagP}
\ee
This fact was first pointed out in \cite{Baulieu:2001fi} where it was also shown to agree with the relation between open string data $(g,\Pi)$ and closed string data $(G,B)$ \cite{Seiberg:1999vs}.

We are primarily concerned with the Hamiltonian formulation since we are set to derive an explicit
expression for the worldsheet current algebra. In full generality it is conceivable to have both a two form $B$-field and a two vector $\Pi$, the resulting Hamiltonian would be
\bea
\cH_{B,\Pi}&=& g^{ij} (p_i+B_{ik}\del x^k) (p_j + B_{jl} \del x^l ) \non \\
&& + g_{kl} (\del x^k + \Pi^{ki}(p_i+B_{im}\del x^m)  )(\del x^l + \Pi^{lj}  (p_j + B_{jn} \del x^n )) . \label{HBPi}
\eea
We can perform the same change of co-ordinates \eq{ptransform}, obtain the same Hamiltonian in a compressed form
\be
\cH_{B,\Pi}= g^{ij} p_i p_j + g_{kl} (\del x^k + \Pi^{ki}p_i )(\del x^l + \Pi^{lj} p_j)   \label{HBPi2}
\ee
and then the Poisson brackets are twisted as in \eq{twpoisson}.

%%%%%%%%%%%%%%%%%%%%%%%%%%%%%%%%%%%%

\subsection{The Charge Algebra}

We are now ready to compute the current algebra associated to \eq{HBPi2}. The most general worldsheet current is
\be
J_{(u,\alpha)}(\sig) = u^i p_i (\sig)+ \alpha_i(\del x^i + \Pi^{ik} p_k) (\sig)
\ee
where $(u^i,\alpha_i)$ are in general allowed to be functions of the $x^j$. We will address the issue of whether this current is conserved in the next section, in the meantime we are free to compute the algebra. It is convenient to perform this calculation in three parts, the bracket of  a vector-vector, a vector-one form and a one form-one form.

%%%%%%%%%%%%%%%%%%%%%%%%%%%%%%%%%%%%
\subsubsection{Vector-Vector Bracket}

We first compute the vector-vector bracket,
\bea
&&\int d\sig d\sig' \{ u^i p(\sig)_i, v^j p_j(\sig') \} \non \\
&&=-\int d\sig \Blp [u,v]^i p_i +  u^i v^j H_{ijk}\del x^k \Brp (\sig)\non \\
&&=-\int d\sig\Blp \blp [u,v]^l -  u^{[i} v^{j]} H_{ijk} \Pi^{kl}\brp p_l +  u^{[i} v^{j]}H_{ijk}(\del x^k + \Pi^{kl}p_l)\Brp (\sig) .
\eea
We see here that the Lie bracket on vector fields has been deformed in the combined presence of a bi-vector and an $H$-field while the $H$-coefficient is insensitive to the bi-vector.

%%%%%%%%%%%%%%%%%%%%%%%%%%%%%%%%%%%%

\subsubsection{Vector-Form Bracket}
Here we calculate the term 
\be
\int d\sig d\sig' \{ \alpha_j(\del x^j + \Pi^{ji} p_i)(\sig), v^k p_k(\sig') \}. \non
\ee
One piece is familiar from the Courant bracket 
\bea
&&\int d\sig d\sig'\{ \alpha_j\del x^j(\sig) , v^k p_k(\sig') \}\non \\
 &&=\int d\sig \blp \cL_{\bv}\alpha - \half d \iota_v \alpha \brp_i \del x^i(\sig) .\non 
\eea
The extra term we need is 
\be
\int d\sig d\sig' \{\alpha_j \Pi^{ji} p_i(\sig) , v^k p_k(\sig') \} = \int d\sig \Blp [\bv,\alpha  \Pi]^ip_i  -\alpha_j \Pi^{ji} v^k H_{ikl} \del x^l\Brp 
\ee
where $\alpha \Pi$ is the vector field $\alpha_j  \Pi^{ji} \del_i$ and $[.,.]$ is the usual Lie bracket on vector fields.
and so in total we get
\bea
&& \int d\sig d\sig' \{ \alpha_j(\del x^j + \Pi^{ji} p_i), v^k p_k \} \non \\
&&=\int d\sig  \Blp ([\bv,\alpha  \Pi]^i - (\cL_{\bv}\alpha - \half d \iota_v \alpha)_k \Pi^{kl} +\alpha_m v^i H_{ijk}  \Pi^{jm}  \Pi^{kl} )p_l \non \\
&&+\blp (\cL_{\bv}\alpha - \half d \iota_v \alpha)_l - \alpha_jv^k \Pi^{ji}  H_{ikl}  \brp(\del x^l + \Pi^{lm} p_m) \Brp(\sig).
\eea

%%%%%%%%%%%%%%%%%%%%%%%%%%%%%%%%%%%%

\subsubsection{Form-Form Bracket}

We now calculate the term
\be
\int d\sig d\sig'\{ \alpha_j(\del x^j + \Pi^{ji} p_i),  \beta_k(\del x^k + \Pi^{kl} p_l) \} \label{ff}
\ee
which has the interpretation of a Poisson bracket of two one-forms. 
This term is entirely new since without the twisting by the bi-vector $\Pi^{ij}$, we have 
\be
\{ \alpha_j\del x^j ,  \beta_k \del x^k \}=0\non
\ee 
in the Courant bracket.

First we need the following two terms
\bea
&& \int d\sig d\sig' \{\alpha_i \del x^i (\sig) , \beta_j \Pi^{jk} p_k (\sig')  \}  \non \\
&&\ \ \ \ \ = \int d\sig (\beta_j \Pi^{jk}   \del_k \alpha_i) \del x^i(\sig) + \int  d\sig d\sig' \alpha_i (\sig) \beta_j \Pi^{ji} (\sig')  \del_\sig\delta(\sig-\sig') \label{xx1} \\
&& \non \\
&& \int d\sig d\sig'\{ \alpha_j \Pi^{jk} p_k (\sig), \beta_i \del x^i(\sig')\}\non \\
&&\ \ \ \ \  =- \int d\sig (\alpha_j \Pi^{jk}   \del_k \beta_i) \del x^i(\sig) - \int  d\sig d\sig'\,  \alpha_j \Pi^{ji} (\sig)\beta_i (\sig')  \del_{\sig'}(\sig-\sig')  \label{xx2} .
\eea
Since total derivatives on the worldsheet vanish, the sum of these two pieces is equal to the Koszul bracket  (see Appendix \ref{Kos})
\bea
&& \int d\sig \Blp \blp \beta_j  \del_k \alpha_i- \alpha_j   \del_k \beta_i \brp \Pi^{jk}  -(\beta_j \del_i \alpha_k -  \alpha_j \del_i \beta_k )\Pi^{jk} \Brp \del x^i(\sig) \non \\
&&= \int d\sig [\alpha,\beta]_{\Pi,i} \del x^i(\sig).
\eea
The key to finding the Koszul bracket on the nose is once again that we are computing the charge algebra not the current algebra.

For the remaining term in \eq{ff} we find
\be
\int d\sig d\sig' \{\alpha_i \Pi^{ik}p_k(\sig)  ,  \beta_j \Pi^{jl} p_l(\sig') \}=-\int d\sig \Blp [\alpha \Pi,\beta\Pi]^ip_i + \alpha_i\beta_j  \Pi^{ik}\Pi^{jl}  H_{klm} \del x^m \Brp 
\ee
Some rearrangement is needed to combine all the above terms into something respectable, more importantly something which can be compared to the bracket in \cite{roytenberg-2002-61}. One subtle point is that although
we can dispense with (target space) total derivatives when they are contracted with $\del x^i$, when they are contracted with $\Pi^{ij}p_j$ the resulting object is no longer a worldsheet total derivative and so does not vanish.

Combining all these terms together we get
\bea
&&\int d\sig d\sig'\{ \alpha_j(\del x^j + \Pi^{ji} p_i)(\sig),  \beta_k(\del x^k + \Pi^{kl} p_l) (\sig')\} \non \\
&&= \int d\sig \Blp [\alpha,\beta]_{\Pi,i}-\alpha_{j}\beta_k \Pi^{jm} \Pi^{kn} H_{mni} \Brp (\del x^i + \Pi^{il} p_l) (\sig)\non \\
&&  -\int d\sig \, \alpha_i \beta_j \Blp \half [\Pi,\Pi]_S^{ijk} -   \Pi^{il} \Pi^{jn}  \Pi^{km} H_{lnm} \Brp p_k(\sig)
\eea
where $[.,.]_S$ is the Schouten bracket.

%%%%%%%%%%%%%%%%%%%%%%%%%%%%%%%%%%%%

\subsubsection{The Total Bracket}

It is worth writing down the full Roytenberg bracket as calculated in the previous three subsections. For comparison with the math literature this is best done in co-ordinate free notation. We have calculated the charge algebra
\be
\{ \cQ_{(u\alpha)}, Q_{(v,\beta)}\} =-\cQ_{[(u\alpha),(v,\beta)]_R}
\ee
where
\bea
&&[u+\alpha,v+ \beta]_R \non \\
&&=[u,v] -H \Pi (u,v) \non \\
&& + \cL_{u}\beta - \cL_v \alpha -\half d (\iota_u \beta - \iota_v \alpha)  +  \Pi H  (\alpha, v,.) -  \Pi H( \beta,u,.)\non  \\
&&-[v,\alpha \Pi] + [u,\beta \Pi] +(\cL_v \alpha -\cL_{u}\beta  + \half d (\iota_u \beta - \iota_v \alpha))\Pi + \w^2\Pi H (\alpha,.,v) - \w^2\Pi H (\beta,.,u) \non \\
&& - [\alpha,\beta]_{\Pi} + \w^2\Pi H (\alpha,\beta,.)\non \\
&&+ H(u,v,.) + \Blp \half [\Pi,\Pi]_S - \w^3\Pi H\Brp(\alpha,\beta,.)\ . \label{royt}
\eea
The first two lines in this expression correspond to the $f$ coefficients (the first line is a vector, the second a one-form), the next two lines are the $Q$ coefficients (the third line is a vector, the fourth a one-form) and the final line has the $H$ and $R$ coefficients. Of course, this expression reduces to the Courant bracket \eq{Courant} when $\Pi=0$.

%%%%%%%%%%%%%%%%%%%%%%%%%%%%%%%%%%%%

\subsection{Conserved Charges}

It is of interest to know when these charges are conserved, in fact the calculation is really quite straightforward. Locally, one can rewrite the current
\bea
J_{(u,\alpha)}&=&u^ip_i+\alpha_i (\del x^i + \Pi^{ij} p_j) \non \\
&=& u'^ip_i+\alpha_i \del x^i 
\eea
where $u'^i=u^i+\Pi^{ij} p_j$. This is just a calculational trick, as mentioned earlier the twisting by a bi-vector results from a global issue not addressed in the current work. Then as mentioned in section \ref{hamiltonian}, the Hamiltonian $\cH_{B,\Pi}$ can be rewritten as the Polyakov Hamiltonian \eq{PolyakovH} and so the calculation of section \ref{cons1} goes through unchanged, resulting in the conditions
\bea
\cL_{\bu'}(G)=0,\ \ \cL_{\bu'}(B)=d(\alpha-\iota_{\bu'} B)
\eea
which can be expressed as a condition on $(g,\Pi)$ by using \eq{Gg} and \eq{BPi}.

%%%%%%%%%%%%%%%%%%%%%%%%%%%%%%%%%%%%
%%%%%%%%%%%%%%%%%%%%%%%%%%%%%%%%%%%%

\section{$T^3$ with $H$ Flux and T-duality}

In this section we will explicitly work through the various $T$-dual frames of a three torus with $H_3$ flux on it. It is important to realize that as discussed at length in section \ref{dualitygen}, the currents are not always conserved nonetheless we find the correct algebra.
 
%%%%%%%%%%%%%%%%%%%%%%%%%%%%%%%%%%%%

\subsection{$T3H3$} \label{T3H3}

The starting point of this duality sequence is a $T^3$ with $N$ units of $H_3$ flux. The metric and $B$-field are
\bea
ds^2&=&dw^2+dx^2+dy^2 \non \\
B&=&w\, dx\w dy.
\eea
The structure constants are given by
\be
H_{wxy}=\eps_{wxy},\ \ f=Q=R=0.
\ee
%%%%%%%%%%%%%%%%%%%%%%%%%%%%%%%%%%%%

\subsection{Twisted Torus}

The twisted torus is slightly more non-trivial. The metric is 
\be
ds^2= dw^2 + dx^2 + (dy-Nw dx)^2
\ee
and thus the frames are
\be
e^1=dw,\ \ e^2=dx,\ \ e^3=dy-Nw dx.
\ee
The vector fields dual to these frames are
\be
e_1=\del_w, \ \  e_2= \del_x+ Nw \del_y,\ \ e_3=\del_y
\ee
which satisfy the Lie bracket
\be
[e_1,e_2]=Ne_3.
\ee
The rest of the Courant bracket is
\bea
{\rm [}e_1,e^3{\rm ]}_C&=& -N e^2 \\
{\rm [}e_2,e^3{\rm ]}_C&=&Ne^1.
\eea
So we see from this that the structure constants
\be
f^{w}_{\ xy}=\eps^{w}_{\ xy},\ \ \  H=Q=R=0	
\ee
where we have raised indices on $\eps_{abc}$ with the flat frame metric.

%%%%%%%%%%%%%%%%%%%%%%%%%%%%%%%%%%%%

\subsection{Q-space} \label{Qspacesec}
The $Q$-space has the metric and $B$-field 
\bea
ds^2&=&dw^2+\frac{1}{1+N^2 w^2}(dx^2 + dy^2), \non \\
B&=&\frac{Nw}{1+N^2 w^2} dx\w dy. \label{Qspace2}
\eea
However to derive the $Q$-coefficients we need to trade this for a metric and bi-vector:
\bea
ds^2&=&dw^2+dx^2+dy^2 \non \\
\Pi&=&Nw\del_x\w \del_y. \label{QPi}
\eea 
Note that $\Pi$ is Poisson. This bi-vector was first observed in the context of open string data in \cite{Lowe:2003qy} and elaborated on in \cite{Grange:2006es}.

The basis of frames and dual vector fields are 
\bea
&&e^1=dw,\ e^2=dx,\ e^3=dy,\non \\
&&e_1=\del_w,\ e_2=\del_x,\ e_3=\del_y\non
\eea
and the Roytenberg bracket now gives
\bea
\slb e^2,e^3\srb_R&=& \del_w \Pi^{xy} e^1\non \\
&=& Ne^1, \non \\
\slb e_1,e^2\srb_R&=& N e_3, \non \\
\slb e_1,e^3\srb_R&=& -N e_2 \non
\eea
with cyclic permutations and the bracket on all other basis elements vanishes.  This demonstrates that the structure constants are 
\be
Q^{wx}_{\ \ \, y}= \eps^{wx}_{\ \ \, y},\ \ \ H=f=R=0.
\ee

This result is satisfying since we have resolved the following paradox: the $H$-coefficients are zero but there is $H$ flux in the solution \eq{Qspace2}. By trading the $B$-field for a bi-vector we get the desired co-efficients. 
It is interesting that this $H$ flux is not integral when contracted with the frames to give a scalar however when the $B$-field is traded for a bi-vector, we get integral $Q$-coefficients.  This could well be a clue as to how we can impose global constraints on a given background and determine how we choose between a bi-vector and a two form $B$-field. Since locally the bi-vector and $B$-field can be traded for each other this is a serious problem which must be addressed.

Whereas before the trade for the bi-vector we had $O(d,d)$ valued transition functions and thus were in the realm of {\it non-geometricity}, after the trade we appear to have a perfectly fine manifold.  The bi-vector shifts as $w\ra w+1$ but the structure constants are invariant. So while before, we were wary that the compactification appeared to have regions of string scale curvature, in the bi-vector formalism the space appears to be at large volume everywhere. Alternatively, if one trades the $B$-field of section \ref{T3H3} for bi-vector, then the resulting solution appears to be non-geometric.

So we have agreement with \cite{Shelton:2005cf} where it was conjectured the $H$ structure constants would be transformed into these $Q$ structure constants. The algebra in question in that work arises in a very different context. In particular it requires RR fields, which are conspicuously absent in the Roytenberg bracket.

We also have agreement with \cite{Mathai:2004qq} where it was demonstrated that this space should be a bundle of non-commutative $T^2$'s fibered over $S^1$. The bi-vector \eq{QPi} has support on the $T^2$ fiber and is non-trivially fibered over the base $S^1$.

%%%%%%%%%%%%%%%%%%%%%%%%%%%%%%%%%%%%

\subsection{R-space}

If one could complete this sequence of T-dualities one would end up with structure constants 
\be
R^{wxy}=\eps^{wxy},\ \ H=f=Q=0.
\ee
This $R$-space is a mysterious object, there is scant evidence that if we could make sense of it that this could subsequently be embedded in string theory at all. Perhaps the most pressing question is whether the space admits a geometric description even locally. A compelling argument was presented in \cite{Shelton:2005cf}  that this would not be the case and goes as follows: the $T3H3$ space cannot support a wrapped 3-brane since the $H$-flux obstructs this. After allegedly performing three $T$-dualities, this statement is converted into the fact that one could not probe the {\it space} with D0-branes. As such one concludes that it admits no local geometric description.

Perhaps the most concrete results available on the $R$-space are by Bouwknegt,  Hannabuss, Mathai and Rosenberg \cite{Bouwknegt:2004ap, Mathai:2004qq, Mathai:2004qc}. They follow a standard mathematical procedure which is to study the space of functions on a space instead of the space itself. Once they understand $T$-duality at the level of the K-theory of these function algebras, they find that the algebra corresponding to the $R$-space is non-associative.

With the results of the current paper in hand, one is led to contemplate whether one can use the bi-vector formalism to understand the $R$-space. It is straightforward to find a bi-vector which gives $R^{wxy}=\eps^{wxy}$,
\be
\Pi^{yw}=Nw,\ \ \Pi^{xw}=1,\ \ \Pi^{xy}=0. \label{RPi}
\ee
This is not sufficient to declare victory since we need that $Q=0$ (it is straightforward to get $H,f=0$). This may or may not be possible, at least at the time of writing this is not clear.

There is one term in the $Q$-coefficients which must be non-zero in order to have nonvanishing $R$, namely we need
 \be
 \alpha_i \beta_j \del_k \Pi^{ij} \neq 0,
 \ee 
it seems possible for this to happen and still have $[\alpha, \beta]_\Pi=0$ as long as the one forms $\alpha$ and $\beta$ are not closed. 

One potential obstacle to finding the $R$-space in this manner is the following: if we set $N=0$ the third $T$ dual of a flat torus is again a flat torus. So we expect that if we can write a local description for the $R$-space that to zero-th order in $N$ it is just  a flat metric, this is certainly true for the twisted torus and the $Q$-space . However that $R$ structure constants are quadratic in $\Pi$ (when $H=0$) and thus if $\Pi\sim N$ then it would seem that we have the scaling $R\sim N^2$ at least. The bi-vector \eq{RPi} violates this observation but of course doesn't give $Q=0$ anyways.

Nonetheless if one could find a bi-vector which gave the correct structure constants one would then claim to have agreement with \cite{Mathai:2004qc} since having non-vanishing $R$ structure constants is the presence of an obstruction to the Jacobi-identity being satisfied. In a loose sense, this is tantamount to the space being {\it non-associative}. To understand the $R$-space it is probably necessary to understand how to further couple a three vector into the Roytenberg bracket. Whilst mathematically this is a natural operation \cite{roytenberg-2002-61}, physically it is hard to see how to incorporate this into the string sigma model.

%%%%%%%%%%%%%%%%%%%%%%%%%%%%%%%%%%%%

\subsection{Why Are The Currents Not Conserved?}

The problem encountered in the above example is that while certain currents are conserved before duality, there are not conserved after duality. This brings into question whether this duality is in fact an exact equivalence. It should be stressed that this problem is not confined to the $Q$-space, it is already present in the twisted torus. The solution we will propose here is that the metric and $B$ field do not appear on equal footing in the Polyakov action. 

The metric and B-field are couplings in the sigma model but as spacetime fields they exhibit certain gauge invariances. The $B$-field can be shifted by any closed two-form and the spacetime physics is also invariant under change of co-ordinates. The gauge symmetries of the $B$ field are trivially absorbed by coupling the $B$ field to the worldsheet by a WZ term, since there it is the field strength $H=dB$ which appears. The worldsheet couplings which come from the space time metric however do not appear in any gauge invariant form. This we argue is the reason for the missing duality frame. 

Since the Lie derivative is the generator of diffeomorphisms, the conditions for a current to be conserved \eq{conserved} are that the metric and $B$-field behave in a certain way under gauge transformations of the metric. In the above example, the only time this is satisfied for all three currents is when the sigma model can be formulated such that the parameter $N$ just  appears in a gauge invariant quantity, namely the $T3H3$ space. This hints that if one could formulate the sigma model in terms of gauge invariant quantities then the final duality of the sequence would become manifest. From the current investigation the natural analogue of the $H$ field is the $f,Q,R$. Our conclusion is that if there is a way to reformulate the sigma model more directly in terms of the Roytenberg bracket then all three symmetries would be manifest in each duality frame and thus the $R$ space would be accessible. The resolution might in fact be that since M-theory geometrizes $T$-duality, one should instead look for a membrane action which can be written in terms of the Roytenberg bracket\footnote{\cite{Stojevic:2008qy} contains an attempt in this direction however the author considers a somewhat different deformation of the Courant bracket than considered here and in \cite{roytenberg-2002-61}. Generalizing the work \cite{Roytenberg:2006qz} to include $Q$ asnd $R$ coefficients might be a reasonable strategy}.

%%%%%%%%%%%%%%%%%%%%%%%%%%%%%%%%%%%%
%%%%%%%%%%%%%%%%%%%%%%%%%%%%%%%%%%%%
\section{Future Directions}

The most immediate issue in need of attention is a global understanding of the presence of the bi-vector. As already mentioned, locally the $B$-field and the bi-vector are interchangeable, there must be global conditions which determine which is present. The $Q$-space example might give a clue as to how this works \cite{Halmagyi1}. Since $(g_{ij},\Pi^{ij})$ also appear in the study of open strings \cite{Seiberg:1999vs}, it is important to understand if they are truly open string couplings in the current setting.

It would be also interesting to develop a way to study supergravity reductions in the presence of a bi-vector. The Courant bracket appears as the gauge algebra of the four dimensional gauged supergravity obtained from reduction on a twisted torus with $H$-flux \cite{Kaloper:1999yr}. It would be interesting if we could somehow get the full Roytenberg bracket in this manner. This idea has been considered somewhat in reverse in \cite{Dabholkar:2005ve} however since we now understand the physical origin of the explicit expressions for the Roytenberg bracket, this should be readdressed. 

It is highly desirable to obtain a version of the Gukov-Vafa-Witten superpotential \cite{Gukov:1999ya} for bi-vector backgrounds. As mentioned in section \ref{Qspacesec}, the Q-space appears to be a (potentially) large volume manifold with a bivector background, as such the gravity approximation is fine. It would then be possible to revisit the work \cite{Shelton:2005cf} with more confidence in the effective four dimensional gravity expressions.

One of the motivations for the current work was the study of generalized complex geometries obtained as bi-vector deformations of Calabi-yau geometries \cite{Halmagyi:2007ft, Butti:2007aq}. The examples which can be explicitly written down are the gravity duals to the $\beta$ deformation \cite{Lunin:2005jy} of certain $\N=1$ SYM  theories. In this case the bi-vector which deforms the pure spinors is exactly the bi-vector in the non-commutative product of the field theory and in the correct angular co-ordinates the bi-vector is constant. It has been pointed out that that bi-vector which drives the elusive {\it cubic} deformation of $\N=4$ SYM cannot be  constant in any co-ordinates \cite{Kulaxizi:2006zc}. It would be interesting to understand in which sense the former case is gauge trivial and the latter is gauge non-trivial. In an ideal world understanding this bi-vector better might lead to a solution generating technique which we could use to construct the gravity solution.

There is another interesting approach to studying non-geometric backgrounds developed by Chris Hull and collaborators in what is by now large body of work (see for example \cite{Hull:2004in, Dabholkar:2005ve, Hull:2007zu}). They proceed by doubling the geometry on which the duality group acts. In this way a geometric description is obtained for these {\it non-geometric} backgrounds and this is presumably related in some fashion to the approach taken in this paper. It should be fruitful to work this out exactly.

Another interesting avenue to explore is the addition of RR fluxes. In a very interesting couple of papers, techniques for deforming Calabi-Yau backgrounds by RR fluxes on the worldsheet using the hybrid string \cite{Berkovits:1994wr} have been fleshed out \cite{Linch:2006ig, Linch:2008rw}. One particularly interesting aspect of that work is the parallel with generalized complex geometry, the worldsheet variables are in the form perfectly suited to describe supergravity in the generalized complex geometry variables \cite{Grana:2005sn}. It will be interesting to understand how to combine these ideas with the addition of a bi-vector to the worldsheet theory.

We have presented a computationally useful way of exploring the landscape of non-geometric string vacua in that we have given explicit formulas for the fluxes of mixed indices. There are still many issues left unresolved and exploring them should be quite exciting.

\vskip 5mm
%%%%%%%%%%%%%%%%%%%%%%%%%%%%%%%%%%%%
%%%%%%%%%%%%%%%%%%%%%%%%%%%%%%%%%%%%

\noindent {\bf {\Large Acknowledgments}}\\
It is a pleasure to thank Peter Bouwknegt for bringing my attention to the work of Roytenberg  and for numerous discussions which ultimately led to this project. I would like to sincerely thank Oleg Lunin for crucial challenging conversations over the last few years in particular in relation to the torus example in this paper. I have also benefited from extensive discussions with Ruben Minasian about various global aspects of generalized flux backgrounds. In addition I would like to thank Emil Martinec, Sakura Schafer-Nameki and Alessandro Tomasiello for useful conversations and the Australian National University for hospitality during the initial stages of this project. The author is supported in part by NSF CAREER Grant No. PHY-0094328 and by NSF Grant No. PHY-0401814 and also a Fermi-McCormick fellowship from the University of Chicago.

\begin{appendix}

\section{Conventions}
The signature on the worldsheet is $(-,+)$ and $\eps^{\sig\tau}=+1$. The conventions for antisymmetrization are
\be
u^{[i}v^{j]}= \half(u^i v^j-u^j v^i).
\ee
%%%%%%%%%%%%%%%%%%%%%%%%%%%%%%%%%%%%
\section{The Twisting of the Symplectic Form} \label{Ap1}
As mentioned in the body of the paper, it is convenient to perform a coordinate transformation on  phase space which in turn alters the symplectic form by a shift:
\be
\om= \oint d\sig  \delta x^i\w \delta (p_i - B_{ij} \del x^j) . \non
\ee
We expand this additional term as
\bea
\oint d\sig   \delta x^i\w \delta(B_{ij} \del x^j )&=&\oint d\sig \delta x^i\w {\big (}\del_\sig x^j  \delta(B_{ij})  +B_{ij}  \delta( \del_\sig x^j ) {\big )} \non \\
&=& \oint d\sig  \delta x^i \w\Blp \del_k B_{ij}  \del_\sig  x^j   \delta x^k  - \half \del_k B_{ij} \del_\sig x^k \,  \delta x^j  \Brp \non \\
&=&- \half \oint d\sig\, H_{ijk}  \del x^k  \delta x^i \w \delta x^j ,
\eea
where of course
\be
H_{ijk} = \del_i B_{jk} +\del_j B_{ki} +\del_k B_{ij}. \non
\ee

So in total the twisted symplectic form is
\be
\om=\oint d\sig \Blp   \delta x^i \w \delta p_i + \half H_{ijk} \del x^k \delta x^i \w \delta x^j \Brp
\ee
and from this follows the twisted Poisson brackets \eq{twpoisson}.
%%%%%%%%%%%%%%%%%%%%%%%%%%%%%%%%%%%%

\section{The Koszul Bracket} \label{Kos}

When deriving the Roytenberg bracket in the main text, we need a few details about the Koszul bracket which we now review. A good reference is \cite{kosmannschwarzbach-1995}. The Koszul bracket is less well known than the Lie bracket on vector fields but it is designed to mimick the Lie bracket on the space of differential forms on a Poisson manifold. The co-ordinate independent expression for the Koszul bracket when evaluated on one forms is
\be
[\alpha,\beta]_{\Pi}= \cL_{\Pi \alpha} \beta - \cL_{\Pi\beta} \alpha + d(\Pi(\alpha,\beta)). \label{Koszul}
\ee
Much like how the Schouten bracket is the extension of the Lie bracket to sections of $\w^* T_X$, the Koszul bracket can be extended to arbitrary differential forms (sections of $\w^* T^*_X$).

It is worth unpacking \eq{Koszul} a little bit since all the expressions derived in this paper are manifestly co-ordinate dependent. We have
\bea
[\alpha,\beta]_{\Pi}&=& \iota_{\Pi \alpha} d\beta - \iota_{\Pi \beta} d\alpha + d( \iota_{\Pi \alpha} \beta - \iota_{\Pi \beta} \alpha + \Pi(\alpha,\beta)) \non \\
&=& \Pi^{ij} \alpha_j (\del_i \beta_k - \del_k \beta_i) dx^k -\Pi^{ij} \beta_j (\del_i \alpha_k - \del_k \alpha_i) dx^k \non \\
&&- \del_k (\Pi^{ij} \alpha_i \beta_j) dx^k \non \\
&=&\Blp \Pi^{ij}( \alpha_j \del_i \beta_k  -\beta_j \del_i \alpha_k)  -  \alpha_i \beta_j\del_k  \Pi^{ij} \Brp dx^k \non .
\eea

%%%%%%%%%%%%%%%%%%%%%%%%%%%%%%%%%%%%
%%%%%%%%%%%%%%%%%%%%%%%%%%%%%%%%%%%%

\section{The WZ-Poisson Sigma Model  }

The twisting of one forms by a bi-vector first appeared in the WZ-Poisson sigma model \cite{Klimcik:2001vg}. 
The WZ-Poisson sigma model has the action 
\be
S=\int d^2x \Blp A_i\w dx^i + \half \Pi^{ij} A_i\w A_j + \half B_{ij} dx^i \w dx^j \Brp
\ee
and is obviously a topological theory since the worldsheet metric does not appear in the Lagrangian. If one can integrate out the $A_i$ fields then the theory reduces to the Polyakov action with $g_{ij}\ra 0$.

The equations of motion are
\bea
dx^k&=&-\Pi^{kj} A_j \non \\
d A_i&=& -\half \Blp \del_i \Pi^{jk} A_j \w A_k  + H_{ijk}\, dx^j\w dx^k \Brp
\eea
which is a coupled system of equations and gives
\bea
0&=&d\Blp dx^k+\Pi^{kj} A_j  \Brp \non \\
&=& \del_l \Pi^{kj} d x^l \w A_j - \half \Pi^{kl} \del_l \Pi^{ij} A_{i}\w A_j- \half \Pi^{kl} H_{lmn}dx^m \w dx^n\non \\
&=&-\half \Blp \half [\Pi,\Pi]^{ijk} +\Pi^{il}\Pi^{jm}\Pi^{kn} H_{lmn}  \Brp A_{i} \w A_j
\eea
This can be solved in two ways, either 
\be
\half [\Pi,\Pi]^{ijk} +\Pi^{il}\Pi^{jm}\Pi^{kn} H_{lmn} =0 \label{PH0}
\ee
or else it is a constraint on the $A_i$ field space.

If we set $B=0$ then the WZ-Poisson sigma model reduces to the Poisson sigma model, we have just demonstrated that in this case $\Pi$ actually need not be Poisson for consistency of the theory, it seems that in the literature $\Pi$ is assumed to be Poisson from the get-go. So with $H\neq0$ the same idea holds but  Poisson is replaced by {\it twisted} Poisson.

This is interesting when considered in the context of the worldsheet derivation of the Roytenberg bracket, although constructing the Hamiltonian is somewhat more involved here due to the first order action (thus the momentum define constraints on phase space). The reason for revisiting this point is that the math literature on this topic was largely inspired by \cite{Klimcik:2001vg} and they find that the constraint \eq{PH0} is equivalent to the condition for a Dirac structure. 

If this constraint would always hold, then it would appear that the $R$ coefficients constructed from a bi-vector would always vanish. However we see that even in the WZ-Poisson sigma model these couplings are not necessarily required to satisfy \eq{PH0} at least on a subset of field space. It would be interesting to understand in the non-topological Polyakov model with a bi-vector the meaning of whether \eq{PH0} is satisfied or not.

\end{appendix}

%%%%%%%%%%%%%%%%%%%%%%%%%%%%%%%%%%%%
\bibliographystyle{utphys} \bibliography{RoytPaper8}
%%%%%%%%%%%%%%%%%%%%%%%%%%%%%%%%%%%%

\end{document}